\documentclass{iau}
\usepackage{graphics}
\usepackage{graphicx}
\usepackage{amsmath}
\usepackage{booktabs}
\usepackage{colortbl}
\usepackage{xcolor}
\usepackage{subfig}

\title[Kiloparsec-scale jet-driven feedback in AGN] 
{Kiloparsec-scale jet-driven feedback in AGN probed by highly ionized gas:
a MUSE/VLT perspective}

\author[Rodr\'{\i}guez-Ardila \&
Fonseca-Faria]   
{A. Rodr\'{\i}guez-Ardila$^{1,2}$, M. A. Fonseca-Faria $^2$}

\affiliation{$^1$Laboratório Nacional de Astrofísica, \\ R. dos Estados Unidos, CEP 37504-364, Itajubá - MG, Brazil \\ email:
{\tt aardila@lna.br}
\\[\affilskip]
$^2$Instituto Nacional de Pesquisas Espaciais, \\ Av. dos Astronautas, CEP 12227-010, São José dos Campos - SP, Brazil}

\pubyear{2020}
\volume{359}  
\jname{Galaxy evolution and feedback across different environments}
\editors{T. Storchi-Bergmann, R. Overzier, W. Forman \& R. Riffel, eds.}

\affiliation{$^1$Laboratório Nacional de Astrofísica, \\ R. dos Estados Unidos, CEP 37504-364, Itajubá - MG, Brazil \\ email:
{\tt aardila@lna.br}
\\[\affilskip]
$^2$Instituto Nacional de Pesquisas Espaciais, \\ Av. dos Astronautas, CEP 12227-010, São José dos Campos - SP, Brazil}

\begin{document}

\maketitle

\begin{abstract}
We employ optical spectroscopy from the Multi Unit Spectroscopic Explorer (MUSE) combined with X-ray and radio data to study the highly-ionized gas (HIG) phase of the feedback in a sample of five local nearby Active Galactic Nuclei (AGN). Thanks to the superb field of view and sensitivity of MUSE, we found that the HIG, traced by the coronal line [Fe\,{\sc vii}]~$\lambda$6089, extends to scales not seen before, from 700~pc in Circinus and up to $\sim$2~kpc in NGC~5728 and NGC\,3393. The gas morphology is complex, following closely the radio jet and the X-ray emission. Emission line ratios suggest gas excitation by shocks produced by the passage of the radio jet. This scenario is further supported by the physical conditions derived for the HIG, stressing the importance of the mechanical feedback in AGN with low-power radio jets.

\keywords{active galaxy nuclei, Seyfert, emission lines, spectroscopic.}
\end{abstract}

\firstsection 
\section{Introduction}
Jet-driven outflows are now recognized as an important ingredient in the active galactic nuclei (AGN) feedback scenario. The effects of such a mechanism in low-luminosity radio-quiet AGN are not yet clear. Recent evidences gathered from NIR AO observations
(\cite[Rodr\'{\i}guez-Ardila et al. 2017]{rodri+17}, \cite[May al. 2018]{may+18}, \cite[Jarvis et al. 2019]{harvis+19})
indicate that their impact to the ISM cannot be underestimated. Indeed, studies made on low-power radio sources ($\leq 10^{24}$ W~Hz$^{-1}$), which dominate the radio sky, show that jets can significantly impact their surrounding gaseous medium  (\cite[Sridhar et al. 2020]{Sridhar+19}). This is because low power sources tend to have low velocity jets. As a result, their jets are characterised by a larger component of turbulence compared to jets in powerful radio sources. Therefore, this kinetic channel can be more relevant for galaxy evolution than previously thought (\cite[Wylezalek \& Morganti 2018]{wm18}).

Traditionally, the identification of outflows associated to radio jets in the warm, ionized phase of AGN has been done by means of the [O\,{\sc iii}]~$\lambda$5007 line (\cite[Greene et al. 2011]{greene+11}). However, this line may also carry the contribution from a starburst component, the galaxy disk and the part of the NLR that is not participating in the outflow (\cite[Rodr\'{\i}guez-Ardila \& Fonseca-Faria 2020]{romar+20}). Isolating the different contributions to the line profile is tricky and sometimes subject to large uncertainties due to the lack of enough spectral and angular resolution. 

In this respect, \cite[Rodr\'{\i}guez-Ardila et al. (2006)]{rodri+06} showed that coronal lines (CLs) such as [Fe\,{\sc vii}]~$\lambda$6087 in the optical or [Si\,{\sc vi}]~1.963$\mu$m in the near-infrared (NIR) are excellent tracers of the ionized component of the outflows. The energy required for their production ($\chi \geq 100$~eV, where $\chi$ is the ionization potential required to produce the ion) rules out stellar or galactic origin. Moreover, as shown by \cite[Ferguson et al. (1987)]{fkf+97}, when powered solely by photoionization by the central source in local, low-luminosity AGN ($L_{\rm bol} < 10^{43}$~erg\,s$^{-1}$, where $L_{\rm bol}$ is the bolometric luminosity), the CL emission cannot extend to distances larger than a few hundred of parsecs from the central source. Thus, their detection outside the circumnuclear region is a signature of outflowing gas associated to jet-driven shocks, capable of ionizing gas at large distances and possibly pushing outwards the ionized gas.

With the above in mind, we started a program aimed at: ($i$) studying the role of outflow/jet-induced mode feedback in a local sample of low-luminosity AGN known to have low-power jets and strong coronal line emission; ($ii$) characterize the physical properties of the highest ionized component of the outflow in these sources.

\section{The sample, observations and data analysis}

The sample employed in this work is composed of five local Seyfert~2 AGN ($z~< 0.02$, where $z$ is the redshift), widely known for displaying a low-luminosity radio-jet and a narrow line region (NLR) characterized by a bi-conical structure, previously mapped through the [O\,{\sc iii}]~$\lambda$5007 emission. This is the case of the Circinus Galaxy, NGC\,5728, IC~5063, NGC\,3393 and NGC\,5643. They all display [Fe\,{\sc vii}]~$\lambda$6087 in the nuclear spectrum and a NLR that extends to at least 2~kpc from the AGN. 
Integral Field Unit (IFU) data for the above 5 targets were obtained using MUSE/VLT and
retrieved from the European Southern Observatory science portal. The IFU cube for each source is fully reduced, including calibration in flux (in absolute
units) and wavelength. Details of the observations and
data reduction are provided elsewhere (i.e \cite[Mingozzi et al. 2019]{mingozzi+19}).
 
Each data cube was analyzed making use of a set of custom {\sc python} scripts developed by us as well as
software publicly available in the literature. First, we
rebinned the cube, reducing the total number of spaxels to $\sim$ 10000. We then removed the stellar continuum across the whole spectral range of MUSE  (4700 $-$ 9100~\AA) using {\sc starlight} (\cite[Cid-Fernandes et al. 2005]{cid+05}) and the \cite[Bruzual \& Charlot (2003)]{bc03} stellar libraries. This procedure left us with spectra dominated by the nebular emission, allowing us to focus only on the gas emission properties, free of any continuum contamination. 

Thereafter, we measured the emission line fluxes of H$\alpha$ and H$\beta$ at every spaxel on each galaxy to determine the extinction (Galactic and intrinsic) affecting the gas. This was done assuming an intrinsic line ratio H$\alpha$/H$\beta =$ 3.1 and the \cite[Cardelli et al. (1987)]{ccm87} extinction law. 
All integrated line fluxes measured at each spaxel were corrected by the extinction measured accordingly.

We then constructed maps of the flux distribution for the most important lines such as [O\,{\sc iii}]~$\lambda$5007, H$\beta$, H$\alpha$ and [Fe\,{\sc vii}]~$\lambda$6087. Their inspection revealed that the mid- and high-ionized gas is usually arranged in bi-conical structures, with apex at the AGN. 

In the following sections, we will describe the main results, first for the Circinus galaxy and then for the remainder of the sample. 
 
\section{The case of the Circinus Galaxy}

At the adopted distance of 4.2~Mpc  (1" $\sim$ 20.4 pc), Circinus is the closest Seyfert~2 galaxy to us.  Because of its proximity, angular resolution on scales of a few tens of parsecs can be reached even at seeing-limited conditions. 

Circinus is widely known for its prominent CL spectrum (\cite[Moorwood et al. 1996]{moor+96}, \cite[Rodr\'{\i}guez-Ardila et al. 2006]{rodri+06}). Adaptic optics (AO) observations have revealed that the coronal line region (CLR) in this object extends from the nucleus up to 30~pc (\cite[Prieto et al. 2005]{prieto+05}, \cite[M\"uller-S\'anchez et al. 2006]{muller+06}). Moreover, \cite[Oliva et al. (1999)]{oliva+99}, using optical spectroscopy, reported extended [Fe\,{\sc vii}] emission up to 22'' from the center at a PA=318$^{\rm o}$. This result was recently confirmed by \cite[Mingozzi et al. (2019)]{mingozzi+19}, who found  extended [Fe\,{\sc vii}] emission associated to the ionization cone, but no information about its full extension, morphology and gas physical properties were presented.

\begin{figure}
\begin{centering}
	\includegraphics[scale=0.45]{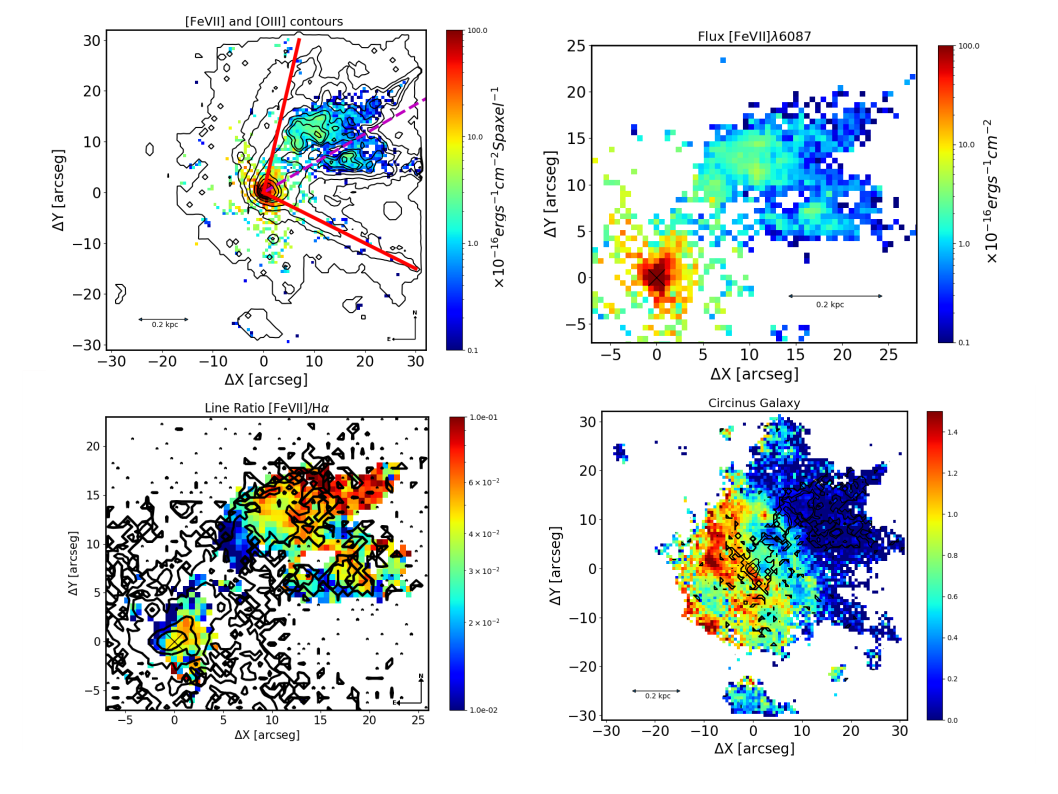}

    \caption{Upper left: Emission map of the [Fe\,{\sc vii}]~$\lambda$6087, overlaid to [O\,{\sc iii}]~$\lambda$5007 contours. The red lines mark the edges of the ionization cone (\cite[Mingozzi et al.  2019]{mingozzi+19}). The dashed magenta line indicates the PA=295$\pm5^{\rm o}$ of the radio continuum reported by \cite[Elmouttie et al. (1998)]{elmo+98}. Upper right: zoom to the NW ionization cone, emphasizing the extended [Fe\,{\sc vii}] emission. Bottom left: Extinction corrected emission line flux ratio [Fe\,{\sc vii}]/H$\alpha$ for the highest ionized portion of the cone. The region in white corresponds to values with signal to noise ratio (S/N) $<$ 3 or where the [Fe\,{\sc vii}] line is not detected. The cross marks the position of the AGN. The contours represent $Chandra$ ACIS image in the 0.5–8 keV band from \cite[Smith \& Wilson (2001)]{sw01}. Bottom right: extinction map ($A_{\rm V}$, in units of mag) of Circinus overlaid to the extended [Fe\,{\sc vii}] emission (black contours).}
    \label{fig:circinus}
    \end{centering}
\end{figure}

The [Fe\,{\sc vii}]~$\lambda$6087 flux distribution constructed for Circinus from MUSE reveals the most extended high-ionization emission ever observed in that AGN. Figure~\ref{fig:circinus} shows that the coronal gas extends up to 700~pc from the central engine. The gas emission appears clumpy, with several knots of emission, concentrated in the innermost region of the ionization cone, following the radio jet axis. The gas displays a complex kinematics (see \cite[Rodr\'{\i}guez-Ardila \& Fonseca-Faria 2020]{romar+20}), implying that it is out of the galaxy plane.  

Figure~\ref{fig:circinus} also shows that the coronal gas is little affected by dust extinction.  Moreover, the spatial coincidence with extended thermal X-ray emission, and its orientation along the radio jet axis, indicates that it is likely the remnant of shells inflated by the passage of the radio jet. The gas velocity dispersion of $\sim$ 300 km\,s$^{-1}$ supports this hypothesis.

The emission line flux ratio [Fe\,{\sc vii}]/H$\alpha$, which directly reflects the ionization state of the gas, is shown in the bottom left panel of Figure~\ref{fig:circinus}. It can be seen that the ratio increases outwards, varying from $\sim$0.01 at $\sim$77~pc from the AGN to $\sim$0.1 in the parcel of gas located at 700~pc from the nucleus. This result cannot be explained by photoionization from the AGN but it is consistent with a scenario where the jet is inducing shocks capable of ionizing gas at large distances and possibly pushing outwards the ionized gas.
 
\section{First detection of kiloparsec-scale [Fe\,{\sc vii}] emission in local AGN}

\begin{figure}
    \begin{centering}
	\includegraphics[scale=0.5]{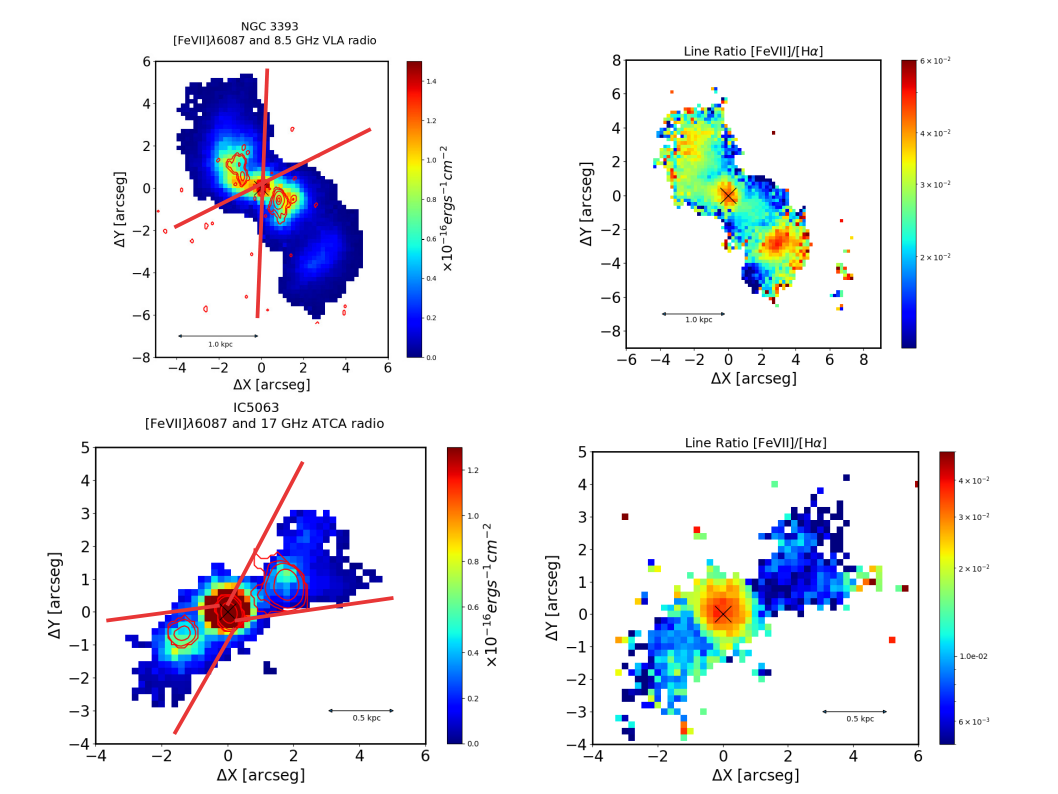}
    \caption{[Fe\,{\sc vii}]~$\lambda$6087 emission line maps for NGC~3393 (upper left) and IC~5063 (bottom left). The full red lines indicate the edges and orientation of the ionization cone. The contours correspond to the radio emission (R. Morganti, {\it private communication}, 2020). The upper and bottom right panels show the line ratio maps (excitation) [Fe\,{\sc vii}]/H$\alpha$ for NGC~3393 and IC\,5063, respectively. White areas correspond to masked values with S/N $<$ 3 or where [Fe\,{\sc vii}] is not detected. In all panels, North is up and East to the left.}
    \label{fig:gals}
    \end{centering}
\end{figure}

A similar analysis as the one described in the preceding section, was applied to NGC\,5728, IC\,5063, NGC\,3393 and NGC\,5643.

Figure~\ref{fig:gals} displays the [Fe\,{\sc vii}] flux distribution for NGC\,3393 and IC\,5063, evidencing the extension of the coronal gas. It can be seen that this energetic emission is arranged within the bi-cones already mapped by means of the [O\,{\sc iii}] line. In NGC\,3393, the [Fe\,{\sc vii}] emission is detected up to 1.3~kpc NE and 2.2~kpc SW from the AGN. Similarly, in IC\,5063, the coronal gas is found up to 1~kpc NW and 0.9~kpc SE. In NGC\,5728, the CL extends up to 2~kpc SE and 1.6~kpc NW (not shown here). Finally, in NGC\,5643, [Fe\,{\sc vii}] is observed up to 0.8~kpc E and 0.9~kpc W to the AGN. To the best of our knowledge, these four Seyfert 2s display the largest CLR already detected in samples of local AGN.   

Figure~\ref{fig:gals} also evidences that the gas distribution is clumpy. Although the CL emission peaks at the AGN position, secondary peaks are also detected along the bi-cones. This is better seen in the upper and bottom right panels of the same figure. They show the excitation map [Fe\,{\sc vii}]/H$\alpha$. In NGC\,3393, in addition to the nucleus, two strong peaks of emission are observed at $\sim$1~kpc NE and SW of the AGN. In IC\,5063, an increase of the gas excitation is evident outside the nuclear region, at $\sim$500~pc NW and SE of the AGN.  

The strong relationship with the radio jet is evident in Figure~\ref{fig:gals}. Radio emission from the jet traces cavities that enhances [Fe\,{\sc vii}]/H$\alpha$, suggesting a role in the excitation of these hot spots. The values of that ratio, between 10$^{-2}$ to 10$^{-1}$, cannot be easily explained by photoionization from the central source. Instead, shock models of \cite[Contini \& Viegas (2001)]{cv01} are able to reproduce the observed gas excitation, along the region where the coronal gas is detected.

\section{Final Remarks}
\begin{itemize}

\item Extended [Fe\,{\sc vii}] emission, at kpc scales, highlights the relevance of the kinetic channel as an important way of depositing energy to the ISM of low-luminosity AGN, even when driven by radiatively poor radio jets.

\item The five sources examined in this work are a showcase of this scenario. We found a conspicuous
[Fe\,{\sc vii}] emission filling the inner cavities of the ionization cone. It extends up to 2~kpc from the AGN in NGC\,3393, NGC\,5728 and IC\,5063.

\item We show that extended coronal emission is associated to the presence of radio jets, lack of dust and extended X-ray emission. The central source plays a fundamental role for gas at distances $R <$ 100-200~pc from the AGN, but at larger $R$, shocks driven by the jet is fundamental to enhance the coronal emission.

\end{itemize}

\end{document}